\documentclass[twocolumn,showpacs,preprintnumbers,amsmath,amssymb]{revtex4}

\usepackage{graphicx}
\usepackage{dcolumn}
\usepackage{bm}

\begin{document}


\title{Maximal violation of the Ardehali's inequality of $n$ qubits}
\author{Zeqian Chen}
\email{zqchen@wipm.ac.cn}
\affiliation{%
Wuhan Institute of Physics and Mathematics, Chinese Academy of
Sciences, P.O.Box 71010, 30 West District, Xiao-Hong Mountain,
Wuhan 430071, China}%

\date{\today}

\begin{abstract}
In this paper, we characterize the maximal violation of Ardehali's
inequality of $n$ qubits by showing that GHZ's states and the
states obtained from them by local unitary transformations are the
unique states that maximally violate the Ardehali's inequalities.
This concludes that Ardehali's inequalities can be used to
characterize maximally entangled states of $n$ qubits, as the same
as Mermin's and Bell-Klyshko's inequalities.
\end{abstract}

\maketitle

As is well known, maximally entangled states, such as Bell states
[1] and GHZ states [2], have become a key concept in the nowadays
quantum mechanics. On the other hand, from a practical point of
view maximally entangled states have found numerous applications
in quantum information [3]. A natural question is then how to
characterize maximally entangled states. There are extensive
earlier works on maximally entangled states [4], however, this
problem is far from being completely understood today. It is
argued [5] that the maximal violation of the Bell-type
inequalities can be used to characterize maximally entangled
states, which are based on Einstein, Podolsky, and Rosen's (EPR's)
notion of local realism [6]. Therefore, for characterizing
maximally entangled states, it is suitable to study the states
that maximally violate the Bell-type inequalities. Recently, the
author [7] has completely resolved the two-qubit case and shown
that the Bell states and the states obtained from them by local
unitary transformations are the unique states that violate
maximally the Clauser-Horne-Shimony-Holt (CHSH) inequality [8].

It is natural to consider the $n$-qubit case in terms of the
Bell-type inequalities. This is not precise, because there are
infinitely many versions of the Bell-type inequalities [9]. It is
well known that Mermin [10] derived an $n$-particle Bell-type
inequality, which provides the first spectacular demonstration of
the fact that there is no limit to the amount by which
quantum-mechanical correlations can exceed the limits imposed by
the premises of EPR's local realism. More recently, the author
[11] has shown that the maximal violation of Mermin's inequalities
characterizes the GHZ states of $n$-qubit as similar to the
two-qubit case in terms of the CHSH inequality, that is, the GHZ
states and the states obtained from them by local unitary
transformations are the unique states that violate maximally
Mermin's inequalities. In this paper, the argument is extended to
Ardehali's inequalities [12], which is an extension of Mermin's
inequalities. We will show that the same statement holds true also
for the Ardehali's inequality of $n$ qubits. Consequently, this
furthermore confirms the fact that from the point of view on EPR's
local realism, the GHZ states are only maximally entangled states
of $n$ qubits, as conjectured by Gisin and Bechmann-Pasquinucci
[5].

Let us give a brief review of the Ardehali's inequality of $n$
qubits. Let $A_j,A'_j$ denote two spin observables on the $j$th
qubit, $j=1,...,n.$ For $n\geq 2,$ define the Bell-Ardehali
operator as following\begin{equation}{\cal A}_n = \Re_{n-1} (A_n -
A'_n) + \Im_{n-1} (A_n +A'_n),\end{equation}where
\begin{equation}\begin{array}{lcl} \Re_n &=&
\Re \left ( \bigotimes^n_{j=1} (A_j + i A'_j ) \right )\\
&=&A_1A_2 \cdots A_n\\&~&- A'_1A'_2A_3 \cdots A_n - \cdots\\&~&+
A'_1A'_2A'_3A'_4A_5 \cdots A_n +\cdots\\&~&-A'_1 \cdots A'_6A_7
\cdots A_n - \cdots\\&~&\cdots,\end{array}\end{equation}and
\begin{equation}\begin{array}{lcl}\Im_n &=&
\Im \left ( \bigotimes^n_{j=1} (A_j + i A'_j ) \right )\\
&=&A'_1A_2 \cdots A_n+ \cdots\\&~&-A'_1A'_2A'_3A_4 \cdots A_n -
\cdots\\&~&+A'_1 \cdots A'_5A_6 \cdots A_n +
\cdots\\&~&\cdots,\end{array}\end{equation}respectively. Here each
line of Eqs.(2) and (3) contains all distinct permutations of the
subscripts that give distinct products. Since the total numbers of
terms in Eqs.(2) and (3) are both $\sum_{j~even} (^n_j)=
\sum_{j~odd} (^n_j)=2^{n-1},$ it concludes that\begin{equation}\|
\Re_n \|, \| \Im_n \| \leq 2^{n-1}.\end{equation}

We now use the following generalized Cirel'son's inequality [13]
to obtain the quantum bound of the Bell-Ardehali operator.

{\it Lemma: Let $A, A'$ denote two observables on the system
$\mathbf{A}$ and $B, B'$ on the system $\mathbf{B}$, respectively.
Define the Bell-CHSH operator on the system $\mathbf{A+B}$ as
following$${\cal B} = A \otimes (B + B') + A' \otimes (B -
B').$$Then,\begin{equation}\| {\cal B}\| \leq 2 \sqrt{2} \max
\{\|A\|, \|A'\| \} \max \{\| B \|, \|B'\| \}.\end{equation}}

By using Eqs.(4) and (5) one has that\begin{equation}\| {\cal A}_n
\| \leq 2 \sqrt{2} \cdot 2^{n-2}\cdot 1 =
2^{n-1/2}.\end{equation}However, assuming ``local realism" [2] one
concludes by using Mermin's elegant technique [10,12] that for $n$
even,\begin{equation}\left | \langle {\cal A}_n \rangle \right |
\leq 2^{n/2},\end{equation} and\begin{equation}\left | \langle
{\cal A}_n \rangle \right | \leq 2^{(n+1)/2},\end{equation}when
$n$ is odd, respectively. The inequalities appearing in Eqs.(7)
and (8) are said to be Ardehali's inequalities. The original
Ardehali's inequality is the case that $A_j = \sigma^j_x$ and
$A'_j = \sigma^j_y$ for $j=1,\ldots, n-1,$ but $A_n =
\sigma^n_{\vec{a}}$ and $A'_n = \sigma^n_{\vec{b}},$ where both
$\vec{a}$ and $\vec{b}$ are in the $x-y$ plane and make
$135^{\circ}$ and $45^{\circ}$ with the $x$ axis respectively.
Clearly, for $n=2$ the Ardehali's inequality reduces to the CHSH
equality. However, for $n \geq 3$ Ardehali [12] showed that the
prediction $2^{n-1/2}$ that quantum mechanics makes for the GHZ
state\begin{equation}|GHZ \rangle = \frac{1}{\sqrt{2}} \left ( |
0\cdots 0 \rangle - | 1\cdots 1 \rangle
\right),\end{equation}maximally violates the original Ardehali's
inequality by an exponentially large factor of $2^{(n-1)/2}$ for
$n$ even or $2^{(n-2)/2}$ for $n$ odd.

The main result we shall prove is that

{\it Theorem: A state $| \varphi \rangle$ of $n$ qubits maximally
violates Eq.$(7)$ for $n$ even or Eq.$(8)$ for $n$ odd, that
is,\begin{equation}\langle \varphi | {\cal A}_n | \varphi \rangle
= 2^{n-1/2},\end{equation}if and only if it can be obtained by a
local unitary transformation of the GHZ state $|GHZ \rangle,$
i.e.,\begin{equation}| \varphi \rangle = U_1 \otimes \cdots
\otimes U_n |GHZ \rangle\end{equation}for some $n$ unitary
operators $U_1,...,U_n$ on ${\bf C}^2.$}

The sufficiency of {\it Theorem} is clear. Indeed, as shown by
Ardehali [12], the GHZ state $|GHZ \rangle$ satisfies Eq.(10) with
$A_j = \sigma^j_x$ and $A'_j = \sigma^j_y$ for $j=1,\ldots, n-1,$
and $A_n = \sigma^n_{\vec{a}}$ and $A'_n = \sigma^n_{\vec{b}},$
where both $\vec{a}$ and $\vec{b}$ are in the $x-y$ plane and make
$135^{\circ}$ and $45^{\circ}$ with the $x$ axis respectively. For
generic states $| \varphi \rangle$ of the form Eq.(11), they
maximally violate Ardehali's inequality of $n$ qubits with the
local unitary transforms of the local spin observables of the
original Ardehali's inequality, i.e., $A_j = U_j \sigma^j_x U^*_j$
and $A'_j = U_j \sigma^j_y U^*_j$ for $j=1,\ldots, n-1,$ and $A_n
= U_n \sigma^n_{\vec{a}}U^*_n$ and $A'_n =U_n
\sigma^n_{\vec{b}}U^*_n,$ where $U^*_j$ is the adjoint operator of
$U_j.$

It remains to prove the necessity. Since\begin{equation}{\cal A}_2
= A_1 (A_2- A'_2) + A'_1 (A_2 + A'_2)\end{equation}coincides with
the Bell-CHSH operator [8], the two-qubit case has been proved by
the author [7]. The key point of our argument involved in [7] is
by using the certain algebraic properties that Pauli's matrices
satisfy [14], which is based on the determination of local spin
observables of the associated Bell operator. The argument has been
extended to the case of Mermin's inequalities by the author [11].
As follows, combining this method and some additional mathematical
techniques we show that every state $| \varphi \rangle$ satisfying
Eq.(10) can be obtained by a local unitary transformation of the
GHZ state $|GHZ \rangle.$

Let us fix some notation. For $A^{(\prime)}_j=
\vec{a}^{(\prime)}_j \cdot \vec{\sigma}_j$ $(1\leq j \leq n),$ we
write$$(A_j,A'_j) = (\vec{a}_j, \vec{a}'_j ), A_j \times A'_j = (
\vec{a}_j \times \vec{a}'_j) \cdot \vec{\sigma}_j.$$Here
$\vec{\sigma}_j$ is the Pauli matrices for the $j$th qubit; the
norms of real vectors $\vec{a}^{(\prime)}_j$ in ${\bf R}^3$ are
equal to $1.$ Clearly,\begin{equation}A_j A'_j = (A_j,A'_j) + i
A_j \times A'_j,\end{equation}\begin{equation} A'_jA_j = (A_j,
A'_j) - i A_j \times A'_j,\end{equation}and\begin{equation}\| A_j
\times A'_j \|^2=1- (A_j, A'_j)^2.\end{equation}We always write
\begin{equation}A''_j = A_j \times A'_j.\end{equation}
By Eqs.(13), (14) and (16), one has that\begin{equation}\left
[A_j,A'_j \right ] = 2i A''_j, \left \{A_j, A'_j \right \} = 2
(A_j, A'_j)\end{equation}where $\left [A_j,A'_j \right ]$ and
$\left \{A_j, A'_j \right \}$ are the commutator and
anticommutator of the spin observables $A_j$ and $A'_j,$
respectively.

As follows, we write $A''_{j_1}A''_{j_2},$ etc., as shorthand for
$$I\otimes \cdots I\otimes A''_{j_1} \otimes I \otimes \cdots I
\otimes A''_{j_2}\otimes I \cdots \otimes I,$$where $I$ is the
identity on a qubit. The proof of {\it Theorem} for necessity
consists of the following four steps:

(i)  At first, we show that if $\| \Re^2_n \| = 2^{2(n-1)}$ for $n
\geq 2,$ then\begin{equation}(A_j,A'_j) =0\end{equation}for all
$j=1,...,n.$

The proof is similar to that of the case of Mermin's inequalities
[11]. Indeed, for $n$ even ($n \geq 2$), the square of the
Bell-type operator Eq.(2) is that\begin{widetext}$$\Re^2_n
=2^{n-1}+ \sum^{n/2}_{k=1} (-1)^k 2^{n-2k-1}\sum_{j_1 <\cdots <
j_{2k}} [A_{j_1},A'_{j_1} ]\cdots [A_{j_{2k}},A'_{j_{2k}} ]+
(-1)^{n/2} \frac{1}{2} \{A_1,A'_1 \}\cdots \{A_n,A'_n \} ,$$where
$\{ j_1, j_2,\ldots, j_{n-2} \}$ is a set of $n-2$ indices each of
which runs from $1$ to $n.$ Hence, by Eq.(17) one has
that\begin{equation}\Re^2_n = 2^{n-1}\left ( 1+ (-1)^{n/2}
(A_1,A'_1)(A_2,A'_2)\cdots (A_n,A'_n)\right ) + 2^{n-1}
\sum^{n/2}_{k=1} \left ( \sum_{1 \leq j_1 < j_2 < \cdots < j_{2k}
\leq n} A''_{j_1} A''_{j_2} \cdots A''_{j_{2k}} \right )
.\end{equation}Then, from Eqs.(15) and (19) it concludes
that\begin{equation}\|\Re^2_n \| \leq 2^{n-1}\left (1 +
(-1)^{n/2}x_1\cdots x_n + \sum^{n/2}_{k=1} \sum_{1 \leq j_1 < j_2
< \cdots < j_{2k} \leq n} \sqrt{(1-x^2_{j_1})(1-x^2_{j_2})\cdots
(1-x^2_{j_{2k}})}\right ),\end{equation}where $x_j =(A_j,A'_j)$
for $j=1,...,n.$ It is easy to check that when $-1 \leq x_j \leq
1$ for $j=1,...,n,$ the function in the right hand of Eq.(20)
attains the maximal value $2^{2(n-1)}$ only at $x_1=\cdots
=x_n=0.$ This concludes Eq.(18).

Analogously, for $n$ odd ($n \geq 3$) one has
that\begin{equation}\Re^2_n = 2^{n-1} \sum^{(n-1)/2}_{k=0} \left (
\sum_{1 \leq j_1 < j_2 < \cdots < j_{2k} \leq n} A''_{j_1}
A''_{j_2} \cdots A''_{j_{2k}} \right
).\end{equation}\end{widetext}Since $\sum^{(n-1)/2}_{k=0}
(^n_{2k})=2^{n-1},$ it immediately concludes from Eqs.(15) and
(21) that if $\| \Re^2_n \| = 2^{2(n-1)}$ for $n \geq 3$ odd, then
Eq.(18) holds true for all $j=1,...,n.$

In fact, repeating the proof of [11] for the Bell-Mermin operator
$\Im_n,$ we can conclude the same result for $\Re_n,$ that is,
{\it each state $| \varphi \rangle$ of $n$ qubits
satisfying\begin{equation}\langle \varphi | \Re_n | \varphi
\rangle = 2^{n-1}\end{equation}if and only if it is of the form
$Eq.(11).$}We omit the details.

(ii) Secondly, we prove that if $\| {\cal A}^2_n \| =
2^{2(n-1/2)}$ for $n \geq 3,$ then\begin{equation}(A_j,A'_j)
=0\end{equation}for all $j=1,...,n.$ In this
case,\begin{widetext}\begin{equation}{\cal A}^2_n = 2 \left (
\Re^2_{n-1} + \Im^2_{n-1} \right ) + 2^n \left ( \sum^m_{k=0}
\sum_{1\leq j_1 < j_2 <\cdots < j_{2k+1} \leq
n-1}A''_{j_1}A''_{j_2}\cdots A''_{j_{2k+1}} \right
)A''_n,\end{equation}where $m = (n-1)/2$ for $n$ odd or $(n-2)/2$
for $n$ even.

Since $A^2_n=A'^2_n =1,$ a simple computation yields
that\begin{equation}{\cal A}^2_n = 2 \left ( 1- (A_n,A'_n) \right
) \Re^2_{n-1} + 2 \left ( 1 + (A_n,A'_n ) \right ) \Im^2_{n-1} -
\left [ \Im_{n-1}, \Re_{n-1} \right ][A_n,A'_n].\end{equation}If
$\| {\cal A}^2_n \| = 2^{2(n-1/2)},$ then by using Eqs.(4) and
(17) one has that $\| \Re^2_{n-1} \| = \| \Im^2_{n-1} \| =
2^{2(n-2)}$ and $[A_n,A'_n]=2.$ Hence, by Step (i) and Eq.(15) one
concludes Eq.(23). On the other hand,$$\left [ \Im_{n-1},
\Re_{n-1} \right ] = \sum^m_{k=0} (-1)^k 2^{n-2k-2} \sum_{1\leq
j_1 < j_2 <\cdots < j_{2k+1} \leq
n-1}[A_{j_1},A'_{j_1}][A_{j_2},A'_{j_2}]\cdots
[A_{j_{2k+1}},A'_{j_{2k+1}}].$$\end{widetext}Therefore, from
Eqs.(17), (23) and (25) one follows Eq.(24).

In particular, if Eq.(23) holds true, it concludes that
\begin{equation}A_j A'_j=-A'_jA_j=iA''_j,\end{equation}
\begin{equation}A'_jA''_j=-A''_jA'_j=iA_j,\end{equation}
\begin{equation}A''_jA_j =-A_jA''_j=iA'_j,\end{equation}
\begin{equation}A^2_j=(A'_j)^2=(A''_j)^2 = 1,\end{equation}
i.e., $\{A_j,A'_j,A''_j\}$ satisfy the algebraic identities that
Pauli's matrices satisfy [14]. Therefore, by choosing
$A''_j$-representation $\{|0 \rangle_j,|1 \rangle_j \},$
i.e.,\begin{equation}A''_j |0 \rangle_j = |0 \rangle_j, A''_j |1
\rangle_j = -|1 \rangle_j,\end{equation}we have
that\begin{equation}A_j |0 \rangle_j =e^{-i \alpha_j} |1
\rangle_j, A_j |1 \rangle_j = e^{i \alpha_j} |0
\rangle_j,\end{equation}
\begin{equation}A'_j |0
\rangle_j = ie^{-i \alpha_j} |1 \rangle_j, A'_j |1 \rangle_j =-i
e^{i \alpha_j} |0 \rangle_j,\end{equation}for some $0\leq \alpha_j
\leq 2 \pi.$ We write $|0\cdot \cdot \cdot 0 \rangle_n,$ etc., as
shorthand for $|0 \rangle_1\otimes \cdot \cdot \cdot \otimes |0
\rangle_n.$ Then, $\{|\epsilon_1 \cdot \cdot \cdot \epsilon_n
\rangle_n : \epsilon_1,..., \epsilon_n = 0,1 \}$ is a orthogonal
basis of the n-qubit system.

(iii) In the third step, we prove that for $n \geq 3,$ a state $|
\varphi \rangle$ of $n$ qubits satisfying\begin{equation} {\cal
A}^2_n | \varphi \rangle = 2^{2(n-1/2)}| \varphi
\rangle\end{equation}must be of the form\begin{equation}| \varphi
\rangle = a| 0\cdot \cdot \cdot 0 \rangle_n + b| 1\cdot \cdot
\cdot 1 \rangle_n,\end{equation}where $|a|^2 + |b|^2 =1.$

Since $\| {\cal A}^2_n \| = 2^{2(n-1/2)},$ by Step (ii) Eq.(23)
holds true and we can uniquely write
$$| \varphi \rangle = \sum_{\epsilon_1,..., \epsilon_n = 0,1}
\lambda_{\epsilon_1 \cdots \epsilon_n }|\epsilon_1 \cdots
\epsilon_n \rangle_n,$$ where $\sum |\lambda_{\epsilon_1 \cdots
\epsilon_n}|^2 = 1.$ In particular, by Eq.(24) we have
that\begin{equation}A''_{j_1} A''_{j_2} \cdots A''_{j_{2k+1}}A''_n
| \varphi \rangle = | \varphi \rangle\end{equation}where $1 \leq
j_1 < j_2 < \cdots < j_{2k+1} \leq n-1.$ By Eq.(30) we conclude
from Eq.(35) that $\lambda_{\epsilon_1 \cdots \epsilon_n} =0$
whenever $\{\epsilon_{j_1}, \epsilon_{j_2},\ldots,
\epsilon_{j_{2k+1}}, \epsilon_n \}$ contains odd number of $1.$
Therefore, besides $\lambda_{0 \cdots 0}$ and $\lambda_{1 \cdots
1},$ one has that $\lambda_{\epsilon_1 \cdots \epsilon_n} =0.$
This concludes Eq.(34) with $a= \lambda_{0 \cdots 0}$ and $b=
\lambda_{1 \cdots 1}.$

(iv) Finally, we prove that if a state $| \varphi \rangle$ of $n$
qubits with $n \geq 3$ satisfies Eq.(10), then
\begin{equation}| \varphi
\rangle = \frac{1}{\sqrt{2}}\left (e^{i \phi} | 0\cdot \cdot \cdot
0 \rangle_n - e^{i \theta} | 1\cdot \cdot \cdot 1 \rangle_n \right
),\end{equation}for some $0 \leq \phi, \theta \leq 2 \pi.$

By Eq.(6), one concludes that Eq.(10) is equivalent to that ${\cal
A}_n |\varphi \rangle = 2^{n-1/2} |\varphi \rangle.$ In this case,
$|\varphi \rangle$ satisfies Eq.(33) and hence is of the form
Eq.(34). Then, by Eq.(1) one has that\begin{widetext}$${\cal A}_n
|\varphi \rangle = a e^{-i \alpha_n}\left [ (1+i) \Im_{n-1} +
(1-i) \Re_{n-1} \right ]|0\cdot \cdot \cdot
0\rangle_{n-1}|1\rangle_n + b e^{i \alpha_n}\left [ (1-i)
\Im_{n-1} + (1+i) \Re_{n-1} \right ]|1\cdot \cdot \cdot
1\rangle_{n-1}|0\rangle_n,$$where $\alpha_n$ appears in Eqs.(31)
and (32) for the $n$th qubit. Consequently,
\begin{equation}b e^{i \alpha_n}
\left [ (1-i) \Im_{n-1} + (1+i) \Re_{n-1} \right ]|1\cdot \cdot
\cdot 1\rangle_{n-1} = 2^{n-1/2}a|0\cdot \cdot \cdot
0\rangle_{n-1} ,\end{equation}\begin{equation}a e^{-i
\alpha_n}\left [ (1+i) \Im_{n-1} + (1-i) \Re_{n-1} \right ]|0\cdot
\cdot \cdot 0\rangle_{n-1}= 2^{n-1/2} b|1\cdot \cdot \cdot
1\rangle_{n-1}.\end{equation}\end{widetext}By Eq.(37) we have
that$$\begin{array}{lcl}2^{n-1/2}|a| & \leq & |b| \left ( |1-i| \|
\Im_{n-1} \| + |1+i| \|\Re_{n-1} \| \right )\\& \leq & |b|\left (
\sqrt{2}\cdot 2^{n-2} + \sqrt{2}\cdot 2^{n-2} \right )\\&=&
|b|2^{n-1/2},\end{array}$$since $\| \Im_{n-1} \|, \| \Re_{n-1}
\|\leq 2^{n-2}$ by Eq.(4). This concludes that $|a| \leq |b|.$
Similarly, by Eq.(38) we have that $|b| \leq |a|$ and so $|a|=|b|=
\frac{1}{\sqrt{2}}.$ Therefore, Eq.(36) holds true.

Now, denote by $V_j$ the unitary transform from the original
$\sigma^j_z$-representation to $A''_j$-representation on the $j$th
qubit, i.e., $V_j |0 \rangle = |0 \rangle_j$ and $V_j |1 \rangle =
|1 \rangle_j,$ and define$$U_1 = V_1 \left ( \begin{array}{ll} e^{
i\phi} & 0
\\~ 0 & 1 \end{array} \right ),U_2 = V_2 \left(
\begin{array}{ll}1 & ~0
\\ 0 & e^{i \theta} \end{array} \right ),$$and $ U_j =V_j$ for
$3 \leq j \leq n.$ Then $U_j$ are all unitary operators on ${\bf
C}^2$ so that Eq.(11) holds true and the proof is complete.

In conclusion, by using some subtle mathematical techniques we
have shown that the maximal violation of Ardehali's inequality of
$n$ qubits only occurs for the states obtained from the GHZ states
by local unitary transformations. This concludes that Ardehali's
inequalities can be used to characterize maximally entangled
states of $n$ qubits, as the same as Bell-Klyshko's inequalities
[7] and Mermin's inequalities [11]. Note that Mermin's
inequalities, Bell-Klyshko's inequalities, and Ardehali's
inequalities are all special instances of all-multipartite
Bell-correlation inequalities for two dichotomic observables per
site derived by Werner-Wolf and \.{Z}ukowski-Brukner [15], and
they are irreducible in the sense of Werner-Wolf [15] that it
cannot be reduced to lower-order Bell-correlation inequalities.
This seems to suggest that the same statement holds true for all
irreducible Bell-correlation inequalities for two dichotomic
observables per site. It is argued that maximally entangled states
should maximally violate Bell-type inequalities [5] and all states
obtained by local unitary transformations of a maximally entangled
state are equally valid entangled states [16], we conclude that
the irreducible Bell-correlation inequalities for two dichotomic
observables per site characterize the GHZ states as maximally
entangled states of $n$ qubits. Finally, we remark that the W
state$$|W \rangle = \frac{1}{\sqrt{n}}\left (|10\cdots 0 \rangle +
|010\cdots 0 \rangle +\cdots + |0\cdots 01 \rangle \right
)$$cannot be obtained from the GHZ states by a local unitary
transformation and hence does not maximally violate some
(irreducible) Bell-correlation inequalities for two dichotomic
observables per site, although it is a ``maximally entangled"
state in the sense described in [17]. Hence, from the point of
view on the violation of EPR's local realism, the W state might
not be regarded as a ``maximally entangled" state and we need some
new ideas for clarity of the W state [18].


\end{document}